\documentclass[%
reprint,
superscriptaddress,
 amsmath,amssymb,
 aps,
prb,
floatfix,
]{revtex4-2}
\usepackage[english]{babel}
\usepackage{comment}
\usepackage{amsmath}
\usepackage{natbib}
\usepackage{ulem}
\usepackage{graphicx}
\usepackage{dcolumn}
\usepackage{bm}
\usepackage{amsmath}
\usepackage{siunitx}
\usepackage{listings}
\usepackage{mathtools}
\usepackage{float}
\usepackage{dsfont}

\usepackage{xcolor}
\usepackage{url}
\usepackage{hyperref}
\hypersetup{
    colorlinks=true,allcolors=blue,}

\begin{document}
\preprint{APS/123-QED}

\title{Nonreciprocal superfluidlike topological spin transport}
\author{Alexey A. Kovalev}
\affiliation{Department of Physics and Astronomy and Nebraska Center for Materials and Nanoscience, University of Nebraska, Lincoln, Nebraska 68588, USA}
\author{Bo Li}
\affiliation{Institute for Advanced Study, Tsinghua University, Beijing, 100084, China}
\author{Edward Schwartz}
\affiliation{Department of Physics and Astronomy and Nebraska Center for Materials and Nanoscience, University of Nebraska, Lincoln, Nebraska 68588, USA}

\date{\today}

\begin{abstract}
We study superfluidlike spin transport facilitated by thermal diffusion of magnetic domain walls, where the positive and negative chiralities of domain walls act as opposite topological charges. The topological charge conservation leads to algebraic decay of spin current carried by domain walls, allowing for the transport of spin over extended distances. We demonstrate that the presence of the Dzyaloshinskii–Moriya interaction can lead to nonreciprocity in spin flow, thus effectively realizing a spin ratchet. In one scenario, the nonreciprocity arises due to diode-like behavior where the nucleation of domain walls is governed by thermal activation for one direction of spin current and by viscous injection for the other direction of spin current.
We confirm our predictions by micromagnetic simulations of domain walls in TmIG nanowire.

\end{abstract}
\pacs{Valid PACS appear here}
\maketitle 
\section{Introduction}
Using spin for transfer of information can make future electronics more energy efficient~\cite{Brataas2020}. Spin currents are used in a type of magnetic memory relying on spin-orbit torque, allowing us to change magnetic state with very little energy loss by employing spin currents~\cite{Han2021}. Recently, spintronics has greatly benefited from using new ideas relying on topology. These ideas give us a mathematical way to discover processes characterized by low dissipation~\cite{Kovalev2018,He2021,Gbel2021}. Studies of topological solitons for spintronic applications have parallels in other fields of physics ranging from metamaterials to black holes~\cite{Bennett1981,PhysRevD.107.084042,Veenstra2024}.

Magnetic insulators~\cite{Brataas2020} and in particular antiferromagnets~\cite{Han2023} are uniquely useful for low dissipation spin transport due to the absence of contributions associated with charge carriers. Long-distance spin superfluid transport relying on the presence of U(1) symmetry has been studied in collinear~\cite{PhysRevLett.112.227201,PhysRevB.90.094408,PhysRevLett.115.237201,PhysRevLett.87.187202,PhysRevB.95.144432,PhysRevLett.116.117201,PhysRevB.96.134434,PhysRevB.103.144412,Stepanov2018,Yuan2018,SciPostPhys,PhysRevResearch.4.023236}
and noncollinear~\cite{PhysRevB.103.L060406,PhysRevB.103.104425} magnets. However, the presence of additional uniaxial anisotropy can break the U(1) symmetry. In this situation, the long-distance spin transport is still possible and it can be carried by topological solitons, such as domain walls (DWs)~\cite{Kosevich1990}, which can be characterized by the conservation of topological charge~\cite{PhysRevB.97.214426}. This can also lead to situations in which magnetic solitons with positive and negative charge can coexist while undergoing Brownian motion at finite temperature~\cite{PhysRev.130.1677,PhysRevB.90.174434,Ivanov1993,PhysRevB.92.020402,PhysRevB.97.214426,PhysRevApplied.16.054002}. In another example of topological solitons, i.e., skyrmions and antiskyrmions with positive and negative topological charge, they can also coexist within the same system~\cite{Hoffmann2017,PhysRevB.104.064417,Goerzen2023}. This situation somewhat resembles semiconductor systems with p- or n-doping where by combining different types of doping one can obtain useful functionality.

\begin{figure}
    \centering
    \includegraphics[width=\columnwidth]{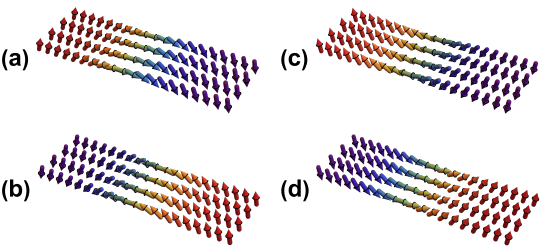}
    \caption{The domain walls with the topological charge $q=1$ in (a) and (b), and the topological charge $q=-1$ in (c) and (d) realizable in an easy-plane magnetic nanowire with additional easy-axis anisotropy.}
    \label{fig:DW}
\end{figure}

\begin{figure}
    \centering
    \includegraphics[width=\columnwidth]{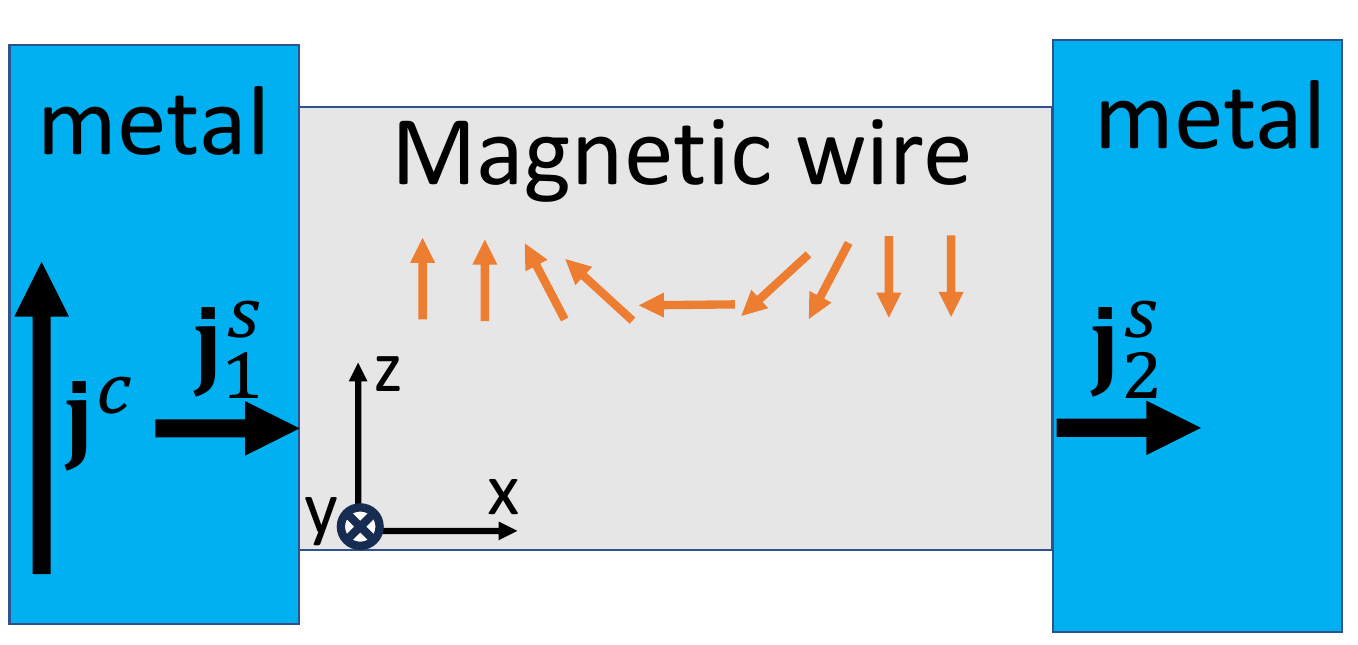}
    \caption{Schematics of a 1D magnetic nanowire that can host domain walls with the topological charge $q=\pm 1$. The nanowire has an easy $x-z$ plane anisotropy with an extra easy $z$-axis. Spin current is injected into the nanowire at the left interface using a heavy metal such as Pt, resulting in preferred injection of domain walls with certain topological charge. Annihilation of domain walls at the right interface generates spin current in the right metal via spin pumping.}
    \label{fig:setup}
\end{figure}

In this work, we study superfluidlike topological spin transport facilitated by thermally populated DWs in an easy-plane ferromagnet with additional in-plane anisotropy. We also include Dzyaloshinskii–Moriya interaction (DMI) which creates preference for one topological charge over the other, effectively realizing topological charge doping. We show how this topological charge doping can lead to nonreciprocity in our system. The notion of topological charge  naturally arises for DWs in the XY ferromagnet and can be associated with the chirality of DWs, see Fig.~\ref{fig:DW} for the types of DWs considered in this work. At low enough temperatures, an easy-plane ferromagnet can effectively approximate the XY ferromagnet. The conservation of topological charge and DW diffusion can then lead to long-distance spin transport with algebraic decay within a typical setup used for observation of spin superfluidity, see Fig.~\ref{fig:setup}.

\section{Domain wall diffusion in a nanowire}
We consider a long ferromagnetic nanowire along the $x$-axis with the cross section $S$ described by the Free energy density:
\begin{eqnarray}
\mathcal U=A(\partial_x\mathbf{n})^2-\kappa n_z^2+K n_y^2+D\hat{y}\cdot(\mathbf n\times \partial_x\mathbf n),\label{eq:energy}
\end{eqnarray}
where  $A$ describes the exchange stiffness, positive $K$ and $\kappa$ correspond to magnetic anisotropies, and $D$ describes the interfacial DMI. We mostly concentrate on the limit $K \gg \kappa$ considered in Ref.~\cite{PhysRevB.92.220409}.
The model in Eq.~\eqref{eq:energy} realizes an easy-plane ferromagnet within $x-z$ plane with an extra easy $z$-axis, thus admitting domain wall (DW) solutions. We describe the local spin density $\mathbf{s}$ using a unit vector $\mathbf{n}$, i.e., $\mathbf{s}=s \mathbf{n}$. The direction is further parametrized using an inplane angle as $\mathbf{n}=(\sin\phi,0,\cos\phi)$.  A static DW solution can be written as
\begin{equation}
\phi(x,X) = \cos^{-1} (q\tanh [\frac{x - X} {\Delta}]) \, ,
\label{eq:DW}
\end{equation}
where $X$ is the position of DW, $q=\pm1$ is the topological charge, and $\Delta=\sqrt{A/\kappa}$ is the DW width. The energy of DW is given by $E^q=S(4\sqrt{A\kappa} +q\pi D)$. The topological charge $q$ can be calculated from the relation
$q =  -\frac{1}{\pi} \int \mathrm{d}x \, \partial_x \phi $
where the integral is taken along the length of the magnetic wire. In the limit of large anisotropy $K$, the DWs are created and annihilated in pairs of opposite charge within the bulk of the magnetic wire. At finite temperatures, a wire with length $L\gg \Delta$ will be characterized by an equilibrium density $\rho_\pm$ of DWs with positive and negative topological charge, where due to finite DMI $\rho_+\neq \rho_-$. Overall, the topological charge described by density, $\rho=\rho_+ - \rho_- $, has to be conserved in the bulk while unpaired charges can be injected through boundaries of the magnetic wire.

The thermal diffusion of DWs at finite temperatures can be described by the  Landau-Lifshitz-Gilbert (LLG) equation:

\begin{equation}
s (1 +  \alpha \mathbf{n} \times) \dot{\mathbf{n}} = \mathbf{n} \times (\mathbf{h} + \mathbf{h}^\text{th}) \, ,
\end{equation}
where $\mathbf{h}^\text{th}$ is the stochastic field described by the correlator $\langle h^\text{th}_i (\mathbf{r}, t) h^\text{th}_j (\mathbf{r}', t') \rangle = 2 \alpha s T \delta_{ij} \delta(\mathbf{r} - \mathbf{r}') \delta(t - t')$ and $\mathbf{h}=- \partial \mathcal U / \partial \mathbf{n}$ is the effective field. We obtain the Langevin equation for the overdamped dynamics of DWs using the collective coordinate approach~\cite{PhysRevLett.100.127204,Kovalev2012,PhysRevB.92.220409} or equivalently the Thiele equation applied to the 1D case for a single variable $X$~\cite{PhysRevLett.30.230,PhysRevB.97.214426},
\begin{equation}
\alpha \eta\dot{X} = F + F^\text{th} \, ,\label{eq:Thiele}
\end{equation}
where $\eta = s \int dV (\partial_x \mathbf{n})^2 = 2 s S / \Delta$ corresponds to dissipative dyadic tensor within the Thiele approach, $F = - \partial U / \partial X$ is the force associated with the field $\mathbf{h}$, and $F^\text{th} = -  \int dV (\mathbf{h}^\text{th} \cdot \partial_x \mathbf{n})$ has the meaning of the stochastic force acting on the DW with the force correlation function, $\langle F^\text{th} (X, t) F^\text{th} (X', t') \rangle = 2 \alpha k_B T \eta \delta(X - X') \delta(t - t')$. Finally, the constant of DW thermal diffusion can be expressed as $\mathcal D = k_B T/\alpha \eta$.
We study the stochastic movement of DWs within a nanowire, giving rise to diffusion characterized by the Fokker-Planck equation for the topological charge and current~\cite{PhysRevB.92.020402},
\begin{equation}
\partial_t \rho + \partial_x I = 0\, ,
\label{eq:fokker}
\end{equation}
where the topological current becomes $I = \mu (F_+\rho_+-F_-\rho_-)- \mathcal D \partial_x \rho$ in the absence of the temperature gradient, and $\mu=1/\alpha\eta$ is the DW mobility. In a steady state, Eq.~\eqref{eq:fokker} realizes long range superfluidlike spin transport due to the conservation of topological charge~\cite{PhysRevB.92.220409}. 

\section{Injection and transport of topological charge} 
To inject a topological current into magnetic nanowire one can use a heavy metal contact in which a charge current $j^c$ induces dampinglike spin-orbit torque, see Fig.~\ref{fig:setup}. The magnetization torque $\boldsymbol{\tau} =\frac{\vartheta j^c}{a} \mathbf{n} \times (\hat{y} \times \mathbf{n}) $ then performs positive or negative work,  
\begin{equation}
W^q	= S \int dt dx \, \boldsymbol{\tau} \cdot (\mathbf{n}\times\partial_t \mathbf{n}) =  q \pi S \vartheta j^c \,,
\end{equation}
during the injection process depending on the sign of charge $q$ where $\vartheta$ is the effective coefficient describing the efficiency of dampinglike spin-orbit torque and $a$ is a small length scale over which the torque is being absorbed~\cite{RevModPhys.91.035004}.

We adopt the reaction-rate theory~\cite{RevModPhys.62.251,PhysRevB.92.220409} to describe the transport of DWs. For each boundary in Fig.~\ref{fig:setup}, we can write the injection rate:
\begin{equation}
I^\pm = \Gamma^\pm (T) - \gamma^\pm (T) \rho^\pm \, , 
\end{equation}
where $\Gamma^\pm (T)$ describes the DW nucleation rate and $\gamma^\pm(T)$ describes the DW annihilation rate per unit density at the boundary. For the nucleation rate, we can write $\Gamma^\pm (T)=\nu(T)\exp(-E^\pm/T+W^\pm/T)$ where $\nu(T)$ is a characteristic frequency describing the nucleation process. For the annihilation rate parametrizing the escape of DWs through the boundary, we identify the topological charge independent $\gamma^0(T)\sim \mathcal D (T)/\Delta$ and the topological charge dependent $\sim q \pi \mu S D/\Delta$ parts.
\begin{figure*}
    \centering
    \includegraphics[width=\textwidth]{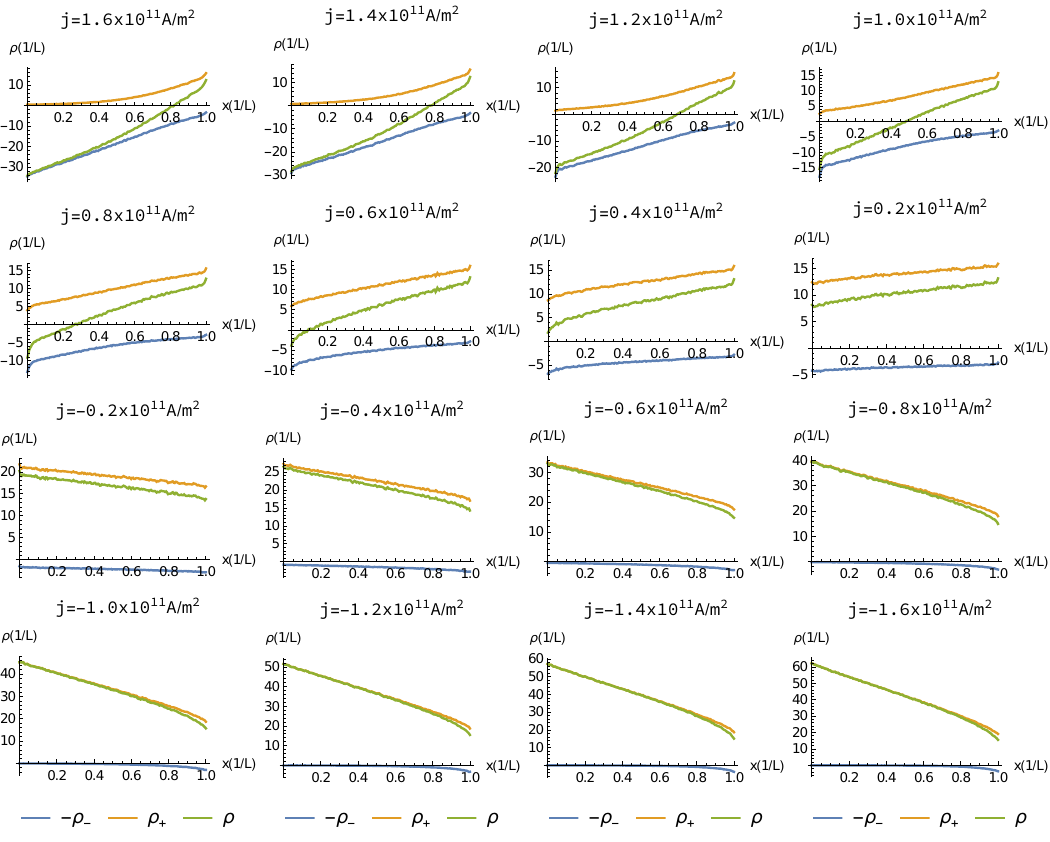}
    \caption{Density of the positive $\rho_+$, negative $\rho_-$, and topological $\rho=\rho_+-\rho_-$ charge as a function of coordinate $x$ for different values of spin-orbit torque $j=\frac{2e}{\hbar}\vartheta j^c$ applied to the left side of the nanowire. }
    \label{fig:density}
\end{figure*}

For injection through the left and right interfaces we can write
\begin{align*} 
 \frac{I^+_L}{\gamma_L^+} - \frac{I^-_L}{\gamma_L^-} &= \rho^+_0 e^{W^+/T}-\rho^-_0 e^{W^-/T} -  \rho_L \, , \\
\frac{I^+_R}{\gamma_R^+} - \frac{I^-_R}{\gamma_R^-} &= -\rho^+_0 +\rho^-_0  +  \rho_R \, ,
\end{align*}
where we use a notation $\rho^\pm_0 =\nu_{L(R)}(T)\exp(-E^\pm/T)/\gamma_{L(R)}^\pm$. 
For simplicity, we initially disregard the topological charge dependent part in $\gamma$ assuming $\gamma_L^+=\gamma_L^-=\gamma^0_L$ and $\gamma_R^+=\gamma_R^-=\gamma^0_R$ and use $I = - \mathcal D \partial_x \rho$ within the magnetic nanowire. This leads to a steady state solution with a uniform topological current:
\begin{equation}
I = \frac{\rho_0^+(e^{W^+/T}-1)-\rho_0^-(e^{W^-/T}-1)}{1/\gamma_L^{0} + 1/\gamma_R^{0} + L/ \mathcal D}  \, . \label{eq:curr}
\end{equation}
Under conditions $\rho_0^+\neq \rho_0^-$ and $|W^q|\sim T$ above equation leads to nonreciprocal spin current. The reciprocity is recovered when $|W^q|\ll T$. Furthermore, at temperatures comparable to $4S\sqrt{AK}/k_B$ the presence of phase slips will invalidate the conservation of topological charge. The numerator in Eq.~\eqref{eq:curr} has the meaning of applied bias while the denominator can be interpreted as the resistance of the system composed of the sum of interfacial resistances $1/\gamma_{L}^{0}$ and $1/\gamma_{R}^{0}$, and the bulk resistance $L/ \mathcal D$. We expect that Eq.~\eqref{eq:curr} will be valid qualitatively in the general case of $\gamma^+ \neq \gamma^-$ as long as the resistance is properly renormalized. As the bias is increased further beyond the values for which the barrier becomes equal to the work performed during the injection, i.e., $W^q=E^q$, the thermally activated behavior is replaced by viscous injection of DWs. The corresponding critical currents are different for different polarities of the bias, 
\begin{equation}
    j^c_\pm=\pm\frac{4\sqrt{A\kappa} \pm \pi D}{\pi \vartheta}.\label{eq:trans}
\end{equation}
This shows that we can realize a diode-like behavior in our system where one direction of spin flow is described by thermally activated behavior while the opposite direction of spin flow is described by viscous injection of DWs. The viscous injection rate can be roughly estimated using the equation of motion of a single domain wall as $I_{inj}=\dot X/\Delta\approx(W^q-E^q)/2\alpha s S \Delta$.
Equation~\eqref{eq:curr} and a possibility of nonreciprocal spin flow carried by topological transport are main results of this paper.

\section{Micromagnetic simulations and material considerations}
To confirm our analytical predictions, we perform micromagnetic simulations using mumax3~\cite{Vansteenkiste2014} code. Within this framework, the thermal field $\mathbf{h}^\text{th}$ with required correlation properties is added to the effective field and the magnetization dynamics is found by numerical integration. The magnetization torque $\boldsymbol{\tau} =\frac{\vartheta j^c}{a} \mathbf{n} \times (\hat{y} \times \mathbf{n}) $ is applied to the leftmost spins of 1D system in Fig.~\ref{fig:setup} where $a$ is the lattice spacing. We first confirm that the simulation has reached the steady state, then we perform averaging over 20000 uncorrelated spin configurations by running different instances and taking snapshots at different times. 

We use parameters corresponding to Tm$_3$Fe$_5$O$_{12}$ (TmIG/Pt) nanowire that can be grown on gadolinium gallium garnet (GGG or SGGG)~\cite{Shao2019,PhysRevB.102.054425}. We consider long thin nanostrip with thickness $t=8$~nm and width $w=40$~nm. For an infinitely long nanowire with elliptical cross section, we expect a shape anisotropy $\frac{\mu_0 M_s^2 t}{2(t+w)}$ along the $x$ axis and an easy $x$-$y$ plane shape anisotropy $\frac{\mu_0 M_s^2 w}{2(t+w)}$.  The (111)-oriented epitaxial iron garnet films also have easy axis anisotropy $K_u$ along the $z$ axis which can be tuned by strain. Overall, such system can be tuned to realize a dominant easy $x$-$z$ plane anosotropy with $K=\frac{\mu_0 M_s^2 t}{2(t+w)}$ and a smaller easy axis anisotropy with $\kappa=K_u-\frac{\mu_0 M_s^2 w}{2(t+w)}-\frac{\mu_0 M_s^2 t}{2(t+w)}$, see Eq.~\eqref{eq:energy}. It is known that for TmIG/Pt thin films grown on GGG or SGGG substrates, the value of an easy axis anisotropy $K_u-\mu_0 M_s^2/2$ is highly tunable by substrate induced strain and composition, and can change between the easy axis and easy plane regimes~\cite{Rosenberg2021}. If not specified otherwise, the simulations are performed at temperature $T=300$K, and we take typical for TmIG/Pt film material parameters: the exchange stiffness $A=1.8\times10^{-12}\,$~J/m, DMI $D=0.015$~mJ/m$^2$, $K=1.1\times10^3$ J/m$^3$, $\kappa=1.1\times10^2$~J/m$^3$, and the Gilbert damping $\alpha=0.01$~\cite{Shao2019,PhysRevB.102.054425}. We simulate a nanowire containing $1600$ sites with lattice spacing $a=8$nm. To characterize the strength of spin-orbit torque, we introduce a parameter $j=\frac{2e}{\hbar}\vartheta j^c$.
\begin{figure}
    \centering
    \includegraphics[width=\columnwidth]{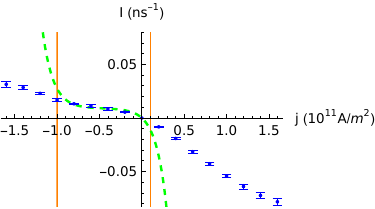}
    \caption{The dots represent the topological current $I$ at the right side of the nanowire as a function of the strength of spin-orbit torque  $j=\frac{2e}{\hbar}\vartheta j^c$ applied to the left side of the nanowire. The plot is obtained using micromagnetic simulations. The dashed line corresponds to fit to Eq.~\eqref{eq:curr}. The vertical lines mark transition to viscous injection mechanism at larger biases.}
    \label{fig:current}
\end{figure}

In Fig.~\ref{fig:density}, we calculate the density of positively and negatively charged DWs, $\rho_\pm$, as well as the topological charge density, $\rho$, for different strengths and signs of injected spin currents in a steady state. The spin current is injected from the left lead, as shown in Fig.~\ref{fig:setup}. We first estimate the currents characterizing the transition from thermally activated behavior to viscous injection in Eq.~\eqref{eq:trans} arriving at $j_+=\frac{2e}{\hbar}\vartheta j^c_+=1.0\times10^{11}$A/m$^2$ and $j_-=\frac{2e}{\hbar}\vartheta j^c_-=-0.1\times10^{11}$A/m$^2$, for the above choice of parameters. This results in diode-like asymmetry of topological flows in Fig.~\ref{fig:density}, which becomes larger as we diminish the Gilbert damping (not shown in the figure).
We also observe that for large biases the slope of topological density curves can depend on densities of positive and negative DWs, especially when one type of DWs becomes depleted. 

To get further insight into the topological transport, we calculate the topological current in Fig.~\ref{fig:current} as a function of spin-orbit torque strength $j$.
We calculate the azimuthal angle and the topological current at the right interface of setup in Fig.~\ref{fig:setup} using the following relation:
\begin{equation}
    I=\dot \phi/\pi.
\end{equation}
We can use the topological current to calculate the spin current density pumped into the right metal using the relation $j^s_R=\hat{y}\cdot[\hbar(g_R^{\prime} + g_R \mathbf{n} \times) \dot{\mathbf{n}}/4\pi]=\hbar g_R I/4$, where we are only interested in the $y$ polarization of spin current (see Fig.~\ref{fig:setup}), and $g_R$ and $g_R^{\prime}$ are the real in imaginary parts of the effective spin mixing conductance of the right interface between the nonmagnetic lead and the magnetic wire~\cite{RevModPhys.77.1375}. In Fig.~\ref{fig:current}, we also plot the topological current calculated using Eq.~\eqref{eq:curr} by dashed line where the denominator is treated as a fitting parameter and all other parameters are the same as in micromagnetics. For spin-orbit torque strengths corresponding to the thermal activation regime (the region between the vertical lines), we observe agreement between analytical results and micromagnetics. By taking the ratio of topological current in Fig.~\ref{fig:current} to the slope of topological density plots in Fig.~\ref{fig:density}, we can estimate the diffusion constant. Using slope instances in Fig.~\ref{fig:density} when both topological charges have comparable densities, we arrive at the diffusion constant $\mathcal D\approx 10^{-4}$m$^2/$s which agrees with the estimate based on the Thiele approach in Eq.~\eqref{eq:Thiele}. The thermal diffusion of a single DW also shows agreement with the Thiele approach as has been demonstrated in Ref.~\cite{PhysRevB.97.214426}. At temperatures that are higher than the magnon's energy gap the diffusion constant will be modified due to interactions between DWs and magnons. 

\section{Conclusions} 
We demonstrated the realization of nonreciprocal topological spin transport in a magnetic nanowire. At lower currents, nonreciprocity arises due to asymmetric injection of DWs with opposite topological charges. As we increase the current, we uncover a diode-like behavior where the injection of domain walls for one direction of the current is governed by
thermal activation while injection for the opposite direction of the current is governed by viscous process without any barrier. The nonreciprocity increases at lower temperatures and can also arise in the bulk of a nanowire, at an interface defined by a step-like change in DMI. As our simulations are performed at finite temperatures, we do not expect that small disorder will alter our conclusions. Strong disorder can introduce a threshold current below which spin transport is not possible due to pinning~\cite{PhysRevApplied.16.054002}. It would be interesting to study the effect of disorder on spin superfluidlike transport in the future, as it may lead to intriguing phenomena such as the ratchet motion of domain walls~\cite{Reichhardt2016}. Our predictions can be tested in TmIG/Pt magnetic nanowires~\cite{Vlez2019,Avci2019}. As the realization of nonreciprocity and diode-like behavior are crucial for electronics functionalities, our results pave the way for electronic devices relying on topological spin currents.

\acknowledgements
We thank Se Kwon Kim for useful discussions.
This work was supported by the U.S. Department of Energy, Office of Science, Basic Energy Sciences, under Award No. DE-SC0021019.

\bibliography{main}
\bibliographystyle{apsrev4-2}
\end{document}